# Topoi of Emergence
## Foundations & Applications[1]

Rainer E. Zimmermann

IAG Philosophische Grundlagenprobleme,
FB 1, UGH, Nora-Platiel-Str.1, D – 34127 Kassel /
Clare Hall, UK – Cambridge CB3 9AL[2] /
Lehrgebiet Philosophie, FB 13 AW, FH,
Lothstr.34, D – 80335 München[3]
e-mail: pd00108@mail.lrz-muenchen.de

Wolfram Voelcker

Voelcker & Freunde, Galerie & Consulting
Oranienburger Str. 2, D – 10178 Berlin[4]
e-mail: wolfram@interflugmedia.com

## Abstract

We discuss aspects of emergence by introducing the concept of negator algebra referring to topos-theoretic results obtained by Trifonov [3] some time ago. As a possible application of this concept we present an example from non-linear economic processes concerning the development of the global crude oil market during the second half of the 20$^{th}$ century.

## Introduction

The description of emergent processes has become very important recently with a view to innovative results achieved in discussing self-organizing processes of various kinds, in particular with respect to the concept of „self-organized criticality" put forward by the Santa Fe school. [4] We present here a short summary of a formal approach which aims at the modelling of emergent processes in terms of a recursive algebra combining aspects of topos theory with (more classical) criteria for self-organizing „chaotic" phenomena. This approach deals mainly with very general problems of emergence claiming its universality with respect to all worldly processes. It should be possible in principle however, to apply this approach to a large variety of fields, and we thus give a relatively limited example by discussing an economic process. In a forthcoming paper [5], it will be shown that a new light can be shed on the approach presented here when taking account of recent developments in quantum gravity. Obviously, a true theory of emergence should be able to prove its validity for a large range of processes, and all of them are nothing but partial structures within a unified whole which we can call our physical world.

---

[1] Revised version of a paper presented to the 7$^{th}$ annual meeting of the Chaos Club Munich, 1997. – The basic idea goes back to earlier work. [1] In the meantime, it has been found out that the recursive aspect of evolution is topic of a book by William Poundstone. [2] As this book is presently not available and out of print, it is still unclear what precise relationship could be established between his idea and the idea presented here.
[2] Permanent addresses.
[3] Present address.
[4] Permanent address.

# 1   Basic Definitions

**1.1** *Self-Composition*: Starting from a basically philosophical point of view [6], namely, how the transition from potentiality to actuality ( = emergence) could be described in a consistent way, which is presently a central topic of theoretical physics [7], the concept of „self-composition" gains a decisive meaning: Be Y a for the time being unspecified, appropriate space, and f: Y → Y a mapping on this space. Then a mapping of the form $f^n(x) = f \circ ... \circ f(x)$ [n times] with $x \in Y$ is called *self-composition* (or self-iteration) of f. (Note that the exponent refers to the explicit number of self-compositions actually being performed rather than of indicating any power of f.) The iteration sequence $\{f^0(x), f(x), f^2(x), ..., f^\omega(x)\}$ characterizes then the ground state ($f^0$), the initial state (f), and all further states up to a final state of a given system, if we assume that f defines a dynamic on Y which formally renders the pair (Y, f) to be some such system.

**1.2** *Chaotic Mappings*: Mappings f of the above mentioned type are called chaotic on Y, if i) f is sensitively dependent on boundary conditions (unpredictability), ii) f is topologically transitive (indecomposability), i.e. if there is for each pair of open sets U, V of Y a positive k such that $f^k(U) \cap V$ is non-zero, iii) the periodical points of f lie dense in Y (nucleus of regularity), i.e. if for P:= $\{f^n(x) = x$ of period n$\}$ and $P \subset Y$, closure(P) = Y.

**1.3** *Ground state*: Let us refer the iteration to the formal ground state of the sequence. Call it $W_0$. Then $W_0 \notin \{W\}$. (Note that formally, the ground state is not actually part of the world!) We call any state $W_k$, k > 0, *world state*, and recursive chaotic mapping between world states, *negation* N. Hence, $\{W\}$ is a suitable space of world states, M say, usually discrete (which might be „averaged out", e.g. by decoherence, into a smooth manifold with a metric of dimension d and signature s). Formally, we obtain the sequence $\{N(W_0) = W_1 = \iota, N(W_1) = W_2 = N(N(W_0)) = N^2(W_0), ..., N(W_{n-1}) = W_n = N^n(W_0) = \omega\}$. This describes the evolution of states of a given world from some initial state (which is the first excitation of the ground state) to a final state.

**1.4** *Global boundary conditions*: Obviously, the initial state (and the final state) have to be specified somewhat in terms of adequate boundary conditions. The problem is that globally, knowledge of boundary conditions is scarce. In particular, the main difficulty is to specify the conditions of the possibility for actually having specific boundary conditions. This refers to the problem of the emergence of a world, in the first place. For a realistic, physical world in the sense of modern cosmology, we might at most choose the „Penrose conditions" for this purpose being represented by $\iota = \Phi(F)$, $\Psi(P) = 0$, and $\Phi(F) = 0$, $\omega = \Psi(P)$, respectively, where $\Phi$ and $\Psi$ are the Ricci and Weyl components of the Riemann curvature tensor from Einstein's theory such that Riemann = Ricci + Weyl = $\Phi(F) + \Psi(P)$, Ricci = Energy-Momentum.

**1.5** *Evolution Equation*: The evolution equation shall describe then the actual unfolding of the ground state. Generally, we can characterize such an evolution as a dynamical system with components representation of the form $(dW_0/ds)^n = N^n(W_0)$. Note that again, the exponent has here a merely combinatorial function: It signifies the initiation of a generic „frequency of cadence" (of a rythmical beat or tact frequency) with which the sucession of negations controls the evolution by acting upon the ground state. Hence, this action is one of multi-contextual kind and surpasses the frame of a merely analytical description. The main reason for this is that the internal logic of the process is not one of the usual Boolean type. This is due to its self-reference which is essentially exhibiting a permanent „self-acting" of the ground state onto itself rather than a true sequence of different excitations. This means that what we observe as an evolutionary process (part of the evolving world in the sense of the aforementioned succession of negations) is nothing but a (restricted) perspective of something we cannot actually oberserve completely (in a somewhat absolutely „true" sense), because it is not part of the world (as we are).

## 2 Formalization of the Basic Dynamics

**2.1** *The category of negators*: We call operators which act as negations on state spaces *negators*. They form a category NEG whose objects are world states and whose morphisms are negations. This category can be visualized very much in „historical" terms as a category of varying sets over an index set: $Set^p$ – in analogy to an argument given by Isham earlier when discussing the complete set of d-consistent history propositions in quantum physics (where d refers to the decoherence function). [8] In this sense it can be shown that for subobjects of NEG there is a suitable subobject classifier such that NEG can be made a topos with an internal logic which secures that the semantics of negators has the structure of a Heyting algebra. Even more: After Trifonov has shown [9] that the paradigm A of an R-xenomorph is Grassmannian (or supersymmetric), if A(F), as category of linear algebras over a partially ordered field, has paradigms as ist objects which are themselves non-trivial Grassmann algebras, it is straightforward to assume (though not yet expli-citly proven) that NEG is an R-xenomorph with a Grassmannian paradigm itself. We come back to that in the appendix.

**2.2** *Formalization of the scheme*: The formalization of this scheme of analysis has been demonstrated for the first time choosing the example of the Keller-Segel scenario in biology [10]: Is E the evolution operator of a given non-linear dynamic, and is N(E) its negation, the a formal cycle can be established such that N(E) initializes the transition from stability to instability of the structures constituting this dynamic, and $N^2(E)$ describes the onset of new stability (of structures). Irreversibility secures that $N^2(E) \neq E$. Such a local cycle of three steps $\{E, N(E), N^2(E)\}$ corresponds to a complete transition from ancient actuality through potentiality to a new actuality. This is what we would think of a model representing the onset of emergence. We call this cycle „sandwich layer". Because we can interpret the collection of sandwich layers describing an evolutionary process both in local as well as in global terms, we can take two different perspectives: on the one hand the local perspective which concentrates on the fine structure of the layers one by one, on the other hand the global perspective which visualizes the totality of several (or all) layers at a time, according to whether or not this is what the research actually being undertaken explicitly asks for.

**2.3** *Concatenation of sandwich structures*: Hence, the evolution of a given world can be visualized then as a concatenation of sandwich structures. Each structure has to be analyzed according to its fine structure. The original form of a sandwich is conserved in the sense that independent of the „strength of magnification" as it is applied to a local fine structure, the basic dynamical scheme will always be reproduced. This is the way in which the fractal pattern of the process is being mirrored, reflecting the fact that their internal logic is essentially of Grassmannian type. Hence, there is a universal structure of mediation of worldly evolution which relates the singularity to the totality (on the microscopic as well as macroscopic levels). Note that the fine structure of this mediation characterizes the meaning of complexity for a given world. The idea is that once an appropriate ground state is being unfolded, the „tact sequence" of the succession of negations acting on states is interpreted as a time parameter by the observer (this fact being part of the epistemological inventory constituting the observer's paradigm). So on the one hand, it is not the irreversibility of processes which determines the unfolding and thus the complexity of a world, but it is rather viceversa: It is the fractal unfolding of the fine structure of sandwiches which *produces* complexity and can be visualized by an observer as irreversibility. This argument runs quite parallel to Stuart Kauffman's idea of a „fourth law" of thermodynamics. [11] On the other hand, we realize that „time" is nothing but a „support parameter" which determines the ordering of observations. But with respect to the picture of visualizing evolution as a permanent „self-negation" of the ground state (which is not actually part of the observer's world), there is no time at all. This conforms with the ongoing discussion about the ab-sence of time as a fundamental entity in physics. [12] Hence, concatenation can be visualized as a purely combinatorial process.

**2.4** *Anticipation and duality*: Mike Manthey has shown [13] that in terms of a pure process view anticipatory systems can be characterized by an explicit *project structure* (which in turn can be interpreted in the sense of existential concepts of productivity and subjectivity) and a *hierarchy* which does not only regulate the levels of abstraction, but also those of the real concretion (of evolution). In particular, it can be shown that there is a „morphic" level of abstraction which corresponds to a *self-reflexion* of the system. Manthey models events (alterations of a system's states) and processes (as sequences of events) in terms of a self-composition of a suitable operator parallel to what we have discussed here so far. The representation of „perception" (i.e. composition of structures from sensorial gain of information) can be expressed in this sense by cohomology operations (of the co-boundary), the representation of „actions" (i.e. de-composition of structures to the purpose of reducing complexity) can be expressed by homology operations (of the boundary). Both of them are mediated dually by what Manthey calls a „twisted isomorphism." In case of self-reflecting systems both of these operations can be visualized as the dual ground structure of the negator model introduced here. [14]

**2.5** *Temporality and algebraic action*: In terms of the action of a system then, *anticipation* can be visualized as a result of the mentioned underlying duality. Insofar, anticipation actually „anticipates" the generic „parameter time" of a system (what we have called s within the frame of the evolution equation discussed above): that is, *duality implies temporality*. (This corresponds in fact, to the generic difference between *monistic ontology* and *dualistic epistemology* common for modern philosophy. There are also deep connections to the philosophical theory of consciousness. [15]) With a view to the evolution equation given above, we can call the upper index n *global age* of a system (or a world). As its *local age*, we introduce the number j of actually emerged sandwich layers referring to a suitable cycle. Hence, we follow here the terminology of Prigogine. [16] Obviously, it is straightforward to assume that n be a discrete superposition of all the j's. This might be also helpful with a view to the problem of quantizing time. [17] Furthermore, it can be shown that if a process is being expressed by a given sequence of states (events) and attributed to the group operations of a Manthey action, then this action can be written as a mixed Clifford product. Because the Grassmann algebra is a subalgebra of the Clifford algebra, as e.g. Freund has shown in some detail for the case of the important algebra of quadratic forms [18], we presently assume that the discussion of an R-xenomorph with a *Clifford paradigm* is of central importance for systems describing worldly evolution as a self-differentiation and/or self-reflexion of its own foundation (ground).

## Appendix

*Xenomorph*: Be F a partially ordered field. An F-*xenomorph* is a category A(F) of linear algebras over F. *Paradigms* of an F-xenomorph are A(F)-objects, *actions* are A(F)-arrows. A paradigm is called *rational*, if the space of motions M(A) is a monoid. We apply here the terminology of Trifonov's: Topoi are introduced in this sense as abstract worlds which represent universes of mathematical discourse whose inhabitants can utilize non-Boolean logics for their argumentation (propositional structures). Contrary to the *sensory space* which mainly deals with the observations of researchers, the space of motions is the set of actions of the researcher. Hence, the paradigm is the set of states of knowledge. In particular, it can be shown that the set of all possible actions of a researcher is a topos whose arrows are those mappings which conserve realizations of the mentioned monoid (of the space of motions). It can also be shown: If A is a rational paradigm and the topos of all possible actions is Boolean (non-Boolean), then the paradigm A is classical (non-classical). For a xenomorph F = R of a generic type of „psychology" of an observer, Trifonov demonstrates that an R-xenomorph implies a classical Einstein paradigm, i.e. a world of dimensionality 4 and signature 2. (The proof goes by quaternion algebra!) Also: If A is a non-trivial Grassmann algebra, then the paradigm is the Grassmannian of an R-xenomorph. Because A has a zero divisor, M(A) cannot be a group. Hence, the logic of a Grassmannian paradigm is always non-Boolean, and mathematics is non-classical.

## 3   Example

From 1960 up to the present time, the evolution of the global crude oil market (for *Arabian light crude oil*, as it is called) can be described in terms of four dynamic equilibria (some of which are attractors) determined by evolution equations which are modelled here with computer-aided numerical methods. The appropriate world state is here the state of the chosen market section. Stable structures of the evolution visualized by a sequence of such states turn out to be critical of either exogeneous type (self-organized [19] or excited by stochastic resonance [20]) or of endogeneous type (induced by parameter change or internal noise [21]). Given the presence of fluctuations, new, in some cases stable, dynamic structures do emerge. In practical terms, the time series of the nominal price as measured in US-Dollars per barrel of crude oil can be structured by characteristic segments [Fig.1]: From September 1960 until January 1971 the posted price remained nominally constant at about 1.8. [22] This can be visualized as the initial state of the system. If projected onto an appropriate phase space representation (price at state t over price at t+1), the situation exhibits a fixed point attractor at 1.8 – reflecting the constancy mentioned. In the late 1960s the market reached a critical stage. [Fig.2] The OPEC countries noticed their increa-sed for capital which could not longer be met by the revenues of oil exports. Hence, the system became unstable. This is the onset of negation with respect to the old price level. In 1971, two price agreements displaced the old fixpoint. In September 1973, for the first time, the increasing demand for crude oil (resulting from a worldwide economic upswing) forced the spot price higher up than the officially posted price. Crude oil was evidently higher valued then ever before, and the OPEC raised gradually the posted price in order to match it with the spotmarket price. This represents the transition from instability to a new stability and completes the first sandwich. The same procedure was repeated shortly afterwards, when the OPEC countries tried to keep pace with inflation. Nevertheless, as can be seen from a suitable phase space representation for the development of the years between 1974 and 1979, the displacement of the fixed point took place in a linear way. [Fig.3] In 1979 then, the price began to fluctuate again as buyers reacted in panic to the news that Iran had halted its oil production, and Saudi Arabia proved unable to compensate this by its own production. As a result, the gap between the price posted by OPEC and the concretely attainable prices on the spotmarket grew rapidly. The situation escalated when other nations won market shares and the OPEC reacted by raising prices again. The well-known positive feedback loop between spotmarket prices and officially posted prices led thus to a new price structure at a level of about 34. This completed the second sandwich. The new price level was being maintained, albeit with clear downward trends, until late 1985. The phase space representation displays this development in terms of an extensively periodic evolution. [Fig.4] Due to the increasing competition with countries winning market shares of their own, the OPEC opted finally for reducing the oil output in order to maintain the high price level. As a result, the revenues dropped by about 100 billion US-Dollars for Saudi Arabia alone. As part of a new price strategy, Saudi Arabia doubled its output again in 1985, forcing other OPEC countries to adopt cartel policies tailored to the marketplace while competing oil-producing countries had to cut back their exports. This completed the third sandwich. Saudi Arabia flooded the market causing the price to fall from 27 in late 1985 to 10 in June 1986. This new state has remained stable until 1997. The phase space representation exhibits clearly now a large attractor basin. [Fig.5] The size of this basin can be estimated from the rise in prices following the Golf War's exogeneous shock effect on the market in 1990/91. Once a trajectory settles down on an attractor, it cannot leave it without the support of external factors. Exogeneous effects such as the speculative demand of 1990/91 in the wake of the loss of Kuwaiti oil production drove the price away from the attractor. But as far as it goes, the trajectory still remains within the attractor basin. Hence, the volatility of the fluctuations increased in clear stages during the period described here. At the fixed point for the years 1960 through 1971, it was by definition zero. Then it increased up to the second attractor and later up to the seond and third. This reflected the increase of the stable phases of the price level at 4, 8.64, and 11, respectively. With the volatility, the combinatorial complexity of the system increases.

Hence, a high value of combinatorial complexity is a characteristic for attractors which are deterministically chaotic.

## Figures [in numerical order]

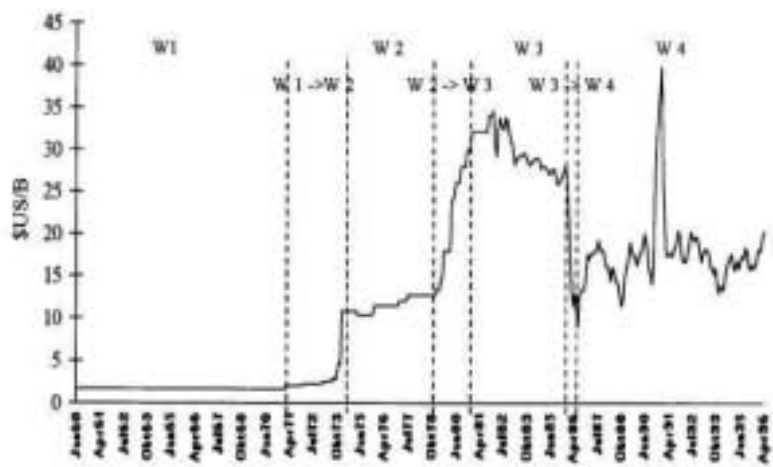

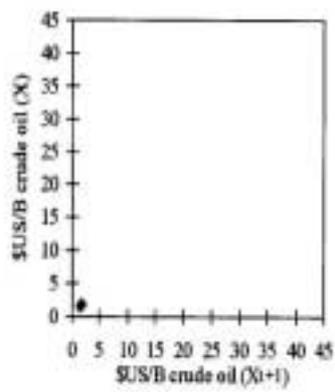

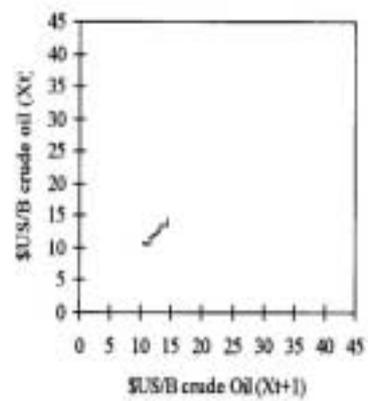

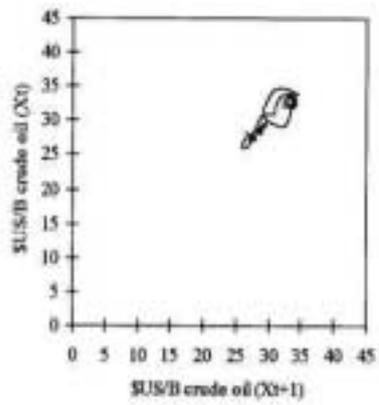
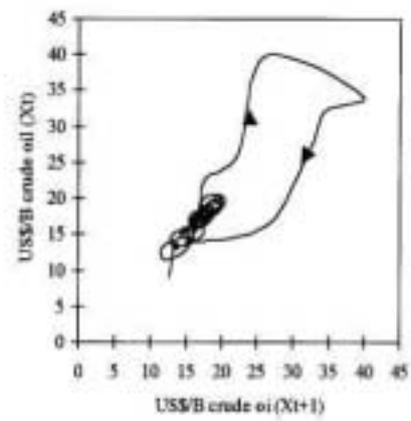